# Coarse-to-Fine Registration of Jawbone CT and Intraoral Scan Data Using GeDi and ICP with Pseudo-IOS Ground Truth

Sho MITARAI*, Hikaru KAYO*, Hisashi OZAKI**, Yuichiro IMAI**, Megumi NAKAO*

* School of Human Health Sciences, Graduate School of Medicine, Kyoto University, Kyoto, Japan.
** Department of Oral and Maxillofacial Surgery, Rakuwakai Otowa Hospital, Kyoto, Japan.

## CT-Pseudo IOS Registration Using GeDi and ICP

### Abstract

In digital dentistry and oral surgery, the registration of jawbone CT and intraoral scanner (IOS) data is essential for integrating internal bone structure with high-resolution dental surface geometry. However, this registration is challenging because the two modalities share only a limited region in common, and their true correspondence is generally unknown. This uncertainty has prevented rigorous quantitative evaluation of registration accuracy. In this study, we propose a pseudo-IOS evaluation framework in which a point cloud emulating an intraoral scan is generated from CT data within the same coordinate frame, so that the transformation between them is known by construction and can serve as a true ground truth. Using this framework, we propose a coarse-to-fine registration method that combines a domain-generalizable local descriptor (GeDi) for initialization-independent global alignment with the iterative closest point (ICP) algorithm for local refinement, and also evaluate the influence of metal artifacts on registration. In experiments on seven cases, ICP alone frequently converged to local minima from a perturbed initial position, whereas GeDi+ICP maintained submillimeter mean absolute error (MAE) across all evaluated jaw and artifact conditions (0.55–0.69 mm). A three-way repeated-measures analysis confirmed that GeDi+ICP was significantly more accurate than GeDi alone. Metal artifacts had a statistically detectable overall effect, but their absolute impact on GeDi+ICP was small.

# 1. Introduction

Three-dimensional imaging integrating hard- and soft-tissue information has become increasingly important for diagnosis and treatment planning in digital dentistry and oral and maxillofacial surgery [1, 2]. Internal bone structure is typically obtained from computed tomography (CT), whereas the precise surface geometry of the dentition is acquired in the form of high-resolution point clouds from an intraoral scanner (IOS). The two modalities are complementary: CT captures osseous structures but suffers from metal artifacts that degrade tooth-crown fidelity [3], whereas IOS reconstructs surfaces accurately but carries no information about underlying bone. Accurate registration of CT and IOS data is therefore essential for surgical design and treatment simulation [2].

The registration of CT and IOS data is made difficult by large differences in resolution, noise, and spatial coverage, with the two modalities sharing only a limited common region around the dentition. Clinically, registration has relied on fiducial markers or manual specification of corresponding points, techniques that are laborious and operator-dependent [1]. Among automatic methods, the iterative closest point (ICP) algorithm [4] and its variants [5] are widely used, but they depend strongly on the initial positions and readily converge to local minima when initialization is poor. To mitigate this, a coarse-to-fine strategy is common, in which a global alignment is first obtained, typically from feature correspondences [6], and is then refined locally.

The accuracy of global alignment depends on establishing reliable correspondences using three-dimensional local descriptors. Manually-defined descriptors, such as fast point feature histograms [7], are computationally light but have limited robustness and cross-domain generalization, whereas learning-based descriptors [8] have recently shown strong performance. Among them, general and distinctive 3D local descriptors (GeDi) [9] is a method that encodes point-cloud patches canonicalized in local reference frames into rotation- and scale-invariant descriptors, and generalizes well across domains not seen during training. This domain generalization is particularly relevant in a correspondence search between heterogeneous modalities such as CT and IOS.

A more fundamental difficulty concerns the evaluation of CT and IOS data alignment because they are acquired separately, and their true correspondence is unknown. Conventional studies have therefore assessed accuracy against manual alignment or surrogate metrics [1], which conflates estimation error with the error of the reference itself. Quantitative evaluation against a true ground truth has thus been essentially unavailable for this task.

This study addresses both issues. First, we propose a *pseudo-IOS* evaluation framework in which a point cloud emulating IOS data is generated from CT data within the same coordinate frame, so that the transformation between them is known by construction and can serve as a true ground truth. Second, we propose a coarse-to-fine registration method that combines GeDi for initialization-independent global alignment with ICP for local refinement. We also examine the influence of metal artifacts, which we simulated in a controlled manner [3], on registration accuracy. The main contributions are: (1) an evaluation framework that enables quantitative assessment of multimodal registration against a true ground truth, and (2) a proposed GeDi+ICP method, whose accuracy and robustness to metal artifacts are demonstrated within this framework.

# 2. Methods

Figure 1 provides an overview of the proposed registration and evaluation framework. The framework consists of two components. First, as shown in Fig. 1(a), GeDi-based correspondence matching provides an initialization-independent global alignment, which is subsequently refined by ICP. Second, paired CT-derived target and pseudo-IOS source clouds (including the simulation of metal artifacts) are constructed (Fig. 1(b)). Because both clouds are generated within the same coordinate system, the ground-truth transformation is known and can be used to quantify registration accuracy. The following sections describe the problem formulation, registration method, and dataset construction in order.

## 2.1 Problem Setting

We address the rigid registration between a pseudo-IOS point cloud and a CT-derived point cloud of the same patient. Both clouds are generated from the same dental CT: the CT-derived dental-region cloud $S_a$ is a solid cloud filled throughout its interior, whereas the pseudo-IOS cloud $P_a^{IOS}$ is an open single-layer surface obtained from $S_a$ by visibility-based extraction, emulating the directly scannable surfaces captured by an IOS. Registration is performed independently for each arch $a \in \{L, U\}$ (mandible L, maxilla U).

The pseudo-IOS cloud $P^{IOS}$ serves as the source, and the CT-derived cloud $S$ as the target. We seek a rigid transformation $g = (R, t)$, with rotation $R \in SO(3)$ and translation $t \in R^3$, to map the source onto the target:

$$g \cdot p = Rp + t,\ p \in P^{IOS}, \tag{1}$$

so that the transformed source $\{Rp + t \mid p \in P^{IOS}\}$ is aligned with $S$.

A key property of this setting is that $P^{IOS}$ is generated from $S$ within the same normalized coordinate frame. The transformation that relates the two clouds is therefore known by construction and serves as the ground-truth transformation $g^*$. In real patients, the true correspondence between the independently acquired CT and IOS is unknown, and the accuracy of an estimated transformation cannot be assessed directly; by contrast, the present setting provides $g^*$, against which an estimated transformation $\hat{g}$ can be quantitatively evaluated.

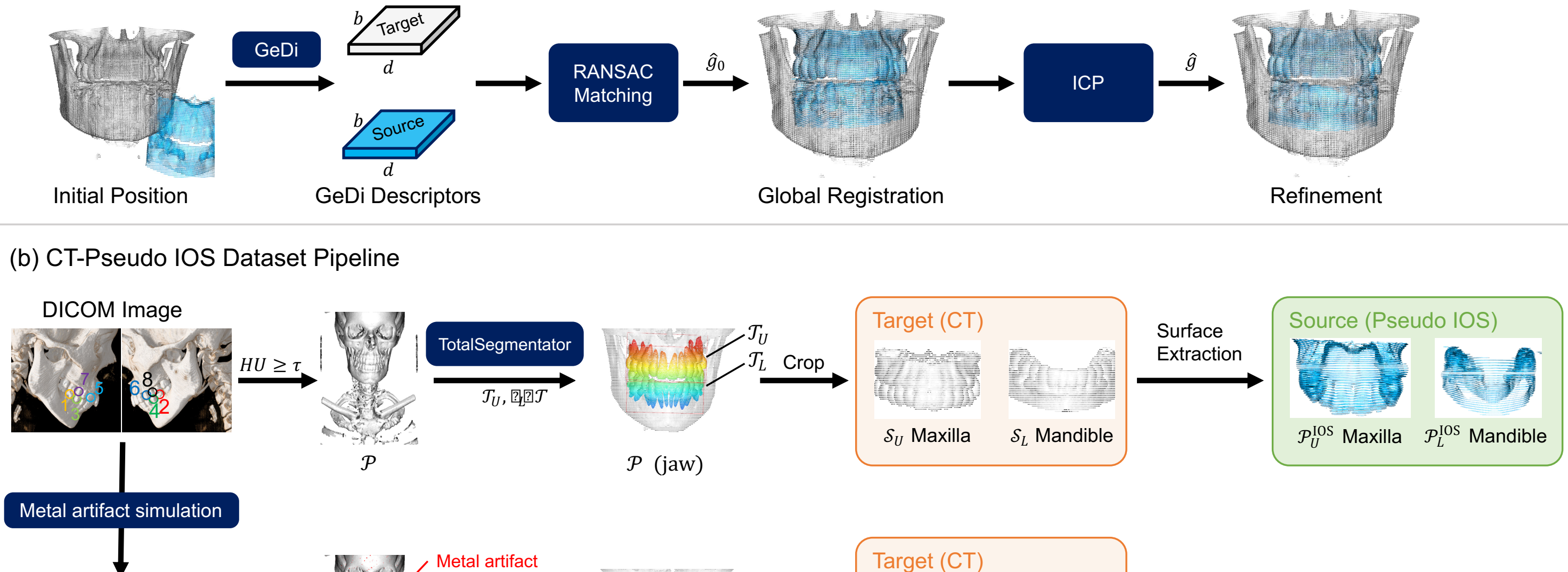

**Fig. 1 Overview of the proposed CT–pseudo-IOS registration and evaluation framework. (a) Coarse-to-fine registration using GeDi-based global alignment followed by ICP refinement. (b) Construction of paired CT-derived target and pseudo-IOS source clouds, including metal-artifact simulation. The pseudo-IOS source generated from the artifact-free CT is paired with either the artifact-free or artifact-affected CT-derived target.**

## 2.2 Coarse-to-Fine Registration using GeDi and ICP

As illustrated in Fig. 1(a), our proposed method estimates the rigid transformation in a coarse-to-fine manner. GeDi first provides an initialization-independent global alignment $\hat{g}_0$ through descriptor-based correspondence matching and random sample consensus (RANSAC) [9]. Using $\hat{g}_0$ as the initial position, ICP then refines the alignment to obtain the final transformation $\hat{g}$.

First, points are randomly sampled for both the source cloud $P^{IOS}$ and the target cloud $S$. At each sampled point, GeDi [10] extracts and canonicalizes a local patch, encoding its multiscale geometry using a PointNet++ backbone [11] to obtain a rotation- and scale-invariant descriptor. Using these per-point descriptors, candidate correspondences are established through nearest-neighbor matching in the descriptor space. Because such matching inevitably produces outliers, especially where the two clouds share only a limited common region or where metal artifacts perturb the local geometry, the rigid transformation is estimated by RANSAC: in each iteration, a minimal subset of correspondences is randomly sampled, a candidate transformation is computed, and the number of inliers within a distance threshold is counted. The transformation supported by the largest number of inliers is then retained and re-estimated from its inliers. This yields the global alignment $\hat{g}_0$, which is independent of the initial position.

Taking $\hat{g}_0$ as the initial position, ICP alternately establishes nearest-point correspondences between the transformed source and the target, updating the rigid transformation that minimizes the distance between corresponding points until the change in registration error becomes sufficiently small. Because ICP is initialized with the descriptor-based global alignment, it avoids the local minima that can arise from a poor initial position and yields the final transformation $\hat{g}$.

## 2.3 CT-Derived Pseudo-IOS Dataset Construction

Figure 1(b) summarizes the construction of the CT–pseudo-IOS dataset. For each artifact-free CT volume, a dental-region point cloud was extracted and normalized and a pseudo-IOS source cloud generated by retaining surfaces visible from virtual scanner viewpoints. Metal artifacts were then simulated to generate artifact-affected target clouds. Because the source and target clouds were derived within the same coordinate frame, their ground-truth transformation and point-level correspondences were known. This procedure was applied independently to the mandible and maxilla, and also independently to each arch $a \in \{L, U\}$ (mandible L, maxilla U).

### 2.3.1 Metal Artifact Simulation

To construct a controlled evaluation setting, we synthesized metal artifacts within artifact-free jawbone CT volumes following the simulation procedure used by Nakao et al. [3]. The purpose of this was to reproduce the streak and dark-band artifacts that dental metals induce in reconstructed CT images, so that the pseudo-IOS registration could be evaluated using data with artifacts resembling those of clinical images.

### 2.3.2 Hard tissue point cloud from CT

As teeth and bone have higher CT values than the surrounding soft tissue, we extract the hard tissue represented by these high values by thresholding the CT values and representing the resulting data as a point cloud. This retaining of voxels above the threshold, and their physical coordinates, yields:

$$\mathcal{P} = \{ (x\,r_x,\ y\,r_y,\ z\,r_z)^\top \mid H(x,y,z) \geq \tau \}, \tag{2}$$

where $H(x,y,z)$ is the CT value at voxel $(x,y,z)$, $(r_x, r_y, r_z)$ are the voxel resolutions (mm) along each axis, and $\tau$ is the threshold for hard-tissue extraction. We set $\tau$ to 300 Hounsfield units (HU) in this study.

Since the target of synthesis is the dentition, we isolate the dental region from $\mathcal{P}$. To localize the teeth, we segment the maxillary and mandibular dentition with TotalSegmentator [12], a deep-learning-based general-purpose CT segmentation method, and use the vertex sets $\mathcal{T}_a$ of the resulting surface meshes. For each arch, an oriented bounding box fitted to $\mathcal{T}_a$ is used to crop $\mathcal{P}$, giving the dental-region cloud $\mathcal{S}_a$ comprising the teeth and the surrounding alveolar bone.

To equalize scale and position across subjects, all clouds are mapped to a common normalized coordinate frame, with each point $\boldsymbol{p}$ transformed as

$$\widetilde{\boldsymbol{p}} = \frac{\boldsymbol{p}-\boldsymbol{\beta}}{\sigma}, \tag{3}$$

where $\boldsymbol{\beta}$ is the center of the axis-aligned bounding box of the CT cloud and $\sigma$ is a scale factor that non-dimensionalizes the coordinates, set to $\sigma = 100$. Hereafter, $\mathcal{S}_a$ refers to the dental-region cloud in this normalized frame.

### 2.3.3 Visibility-based surface extraction

Whereas the CT-derived $\mathcal{S}_a$ is a solid cloud filled throughout its interior, an IOS captures only the outer surface directly visible to the scanner, leaving the interior hollow and the non-scannable deep region (the gingival/alveolar-bone side) absent. We reproduce this open single-layer surface by retaining only the points visible to a virtual scanner. We write the dental region cloud as

$$\mathcal{S}_a = \{\boldsymbol{p}_i\}_{i=1}^{N}, \tag{4}$$

where $\boldsymbol{p}_i \in \mathbb{R}^3$ is the $i$-th point and $N$ is the total number of points. We further introduce a sign $s_a$ for the scanner side: $s_a = +1$ for the mandible, whose occlusal surface faces upward, and $s_a = -1$ for the maxilla, whose occlusal surface faces downward.

The scanning direction is approximately perpendicular to the occlusal plane, but the patient's head position during CT tilts that plane away from the CT body axis. Therefore, we estimate the occlusal axis $\boldsymbol{a}$ by applying principal component analysis to the cloud. Consider the covariance matrix

$$\boldsymbol{\Sigma} = \frac{1}{N}\sum_{i=1}^{N}(\boldsymbol{p}_i - \boldsymbol{\mu})(\boldsymbol{p}_i - \boldsymbol{\mu})^\top, \boldsymbol{\mu} = \frac{1}{N}\sum_{i=1}^{N}\boldsymbol{p}_i, \tag{5}$$

where $\boldsymbol{\mu}$ is the centroid of the cloud. Let $\lambda_1 \leq \lambda_2 \leq \lambda_3$ be the eigenvalues of $\boldsymbol{\Sigma}$ and $\boldsymbol{e}_1, \boldsymbol{e}_2, \boldsymbol{e}_3$ the corresponding eigenvectors. Because the arch is broad within the occlusal plane and thin along the perpendicular, the eigenvector $\boldsymbol{e}_1$ of the smallest eigenvalue $\lambda_1$ approximates the occlusal-plane normal. Orienting it toward the scanner, the occlusal axis is

$$\boldsymbol{a} = \mathrm{sgn}(\boldsymbol{e}_1 \cdot \boldsymbol{z}_s)\,\boldsymbol{e}_1, \tag{6}$$

where $\boldsymbol{z}_s = (0,0,s_a)^\top$ is a reference vector for the approximate scanner direction, " $\cdot$ " denotes the inner product, and $\mathrm{sgn}(\cdot)$ is a sign function.

Reflecting that an IOS scans the arch from multiple directions, we place virtual viewpoints on a hemisphere about the occlusal axis $\boldsymbol{a}$. Taking a set of polar angles $\Theta$, and for each of these $M$ equally spaced azimuth angles, the viewpoint position is

$$\boldsymbol{c}(\theta,\phi) = \boldsymbol{\mu} + \rho\,d\,[\cos\theta\,\boldsymbol{a} + \sin\theta\,(\cos\phi\,\boldsymbol{u} + \sin\phi\,\boldsymbol{v})], \tag{7}$$

where $\theta \in \Theta$ is the polar angle from the occlusal axis, $\phi$ is the azimuth angle, $\boldsymbol{u}$ and $\boldsymbol{v}$ are unit vectors mutually orthogonal to $\boldsymbol{a}$ (so that $\{\boldsymbol{u},\boldsymbol{v},\boldsymbol{a}\}$ is an orthonormal basis), $d$ is the diagonal length of the bounding box of the dental-region cloud, and $\rho$ scales the viewpoint distance relative to $d$. We set $\Theta = \{0°, 30°, 55°\}$, $M = 16$, and $\rho = 1.5$, giving $K = 1 + M(|\Theta| - 1) = 33$ viewpoints in total.

The points visible from each viewpoint are determined with the hidden point removal operator of Katz et al. [13]. For the coordinates $\boldsymbol{q}_i = \boldsymbol{p}_i - \boldsymbol{c}$ centered at a viewpoint $\boldsymbol{c}$, we apply a spherical flipping about a sphere of radius $R$,

$$\widehat{\boldsymbol{q}}_i = \boldsymbol{q}_i + 2\,(R - \|\boldsymbol{q}_i\|)\,\frac{\boldsymbol{q}_i}{\|\boldsymbol{q}_i\|}, \tag{8}$$

where $\widehat{\boldsymbol{q}}_i$ is the flipped point, $\|\cdot\|$ is the Euclidean norm, and $R = \gamma\,d$ is the radius of the flipping sphere, for which a larger factor $\gamma$ classifies more points as visible ($\gamma = 2000$ in this study). A point $\boldsymbol{p}_i$ is deemed visible from $\boldsymbol{c}$ when its flipped image $\widehat{\boldsymbol{q}}_i$ lies on the boundary of the convex hull of $\{\widehat{\boldsymbol{q}}_1, \dots, \widehat{\boldsymbol{q}}_N\}$ together with the origin.

Let $\mathcal{I}_j$ be the index set of points found visible from the $j$-th viewpoint $\boldsymbol{c}_j$. The surface is taken as their union,

$$\mathcal{I} = \bigcup_{j=1}^{K}\mathcal{I}_j, \qquad \mathcal{P}_a^{\mathrm{IOS}} = \{\boldsymbol{p}_i \mid i \in \mathcal{I}\}, \tag{9}$$

where $\mathcal{I}_j$ is the visible-point index set from viewpoint $j$, $\mathcal{I}$ is their union over all viewpoints, and $\mathcal{P}_a^{\mathrm{IOS}}$ is the output point cloud. Through this aggregation, interior points—occluded from every viewpoint—are removed so that the cloud becomes hollow, while the deep region invisible from all viewpoints is left open. Finally, the normals of $\mathcal{P}_a^{\mathrm{IOS}}$ are re-estimated and oriented consistently with the occlusal axis $\boldsymbol{a}$. This step reduces the number of points to roughly $30\%-40\%$ of the solid cloud and yields a hollow, deep-open surface comprising the occlusal, buccal, and lingual aspects.

## 3. Experiment

The purpose of this experiment was to investigate how metal artifacts affect the accuracy of each registration method. Using the pseudo-IOS point clouds and ground-truth transformation defined in Section 2.1, we quantitatively compare descriptor-based, optimization-based, and combined registration approaches, examining the extent to which artifact-induced distortion of the CT-derived point cloud degrades registration performance.

### 3.1 Implementation Details

All experiments were performed on a workstation equipped with an NVIDIA RTX A6000 GPU (48 GB). To evaluate registration under non-trivial initial misalignment, initial misalignments were applied to the source cloud before registration. Initial rotations were generated as random rotation angles within ±10° of each axis, while translations were randomly sampled within ±50 mm along each axis. Each method was required to recover the transformation that realigned the perturbed source cloud with the target. For each case and jaw, registration was repeated 10 times under every artifact condition.

For global registration, 2000 points were randomly sampled from each source and target point cloud. A 32-dimensional GeDi descriptor was computed at each point using the model pretrained on 3DMatch and a local reference frame radius of 0.5 (50 mm). Point normals were estimated using a hybrid k-dimensional-tree search with a radius of 0.01 (1 mm) and a maximum of 30 neighbors. Candidate correspondences were selected by mutual nearest-neighbor descriptor matching. RANSAC estimated a rigid point-to-point transformation without scaling from three correspondences per iteration. The maximum correspondence distance was 0.02 (2 mm), and the edge-length, distance, and normal consistency thresholds were set to 0.9, 0.02 (2 mm), and π/4, respectively. RANSAC was run for up to 50,000 iterations, with a confidence of 1.0.

Local refinement was then performed on the full source and target point clouds using point-to-point ICP with a maximum correspondence distance of 0.1 (10 mm) and without scaling. Each ICP run was initialized with the identity matrix and terminated when the relative changes in both fitness and inlier root mean square error were below $1 \times 10^{-6}$, or after 30 iterations. After each run, the estimated transformation was applied to the current source point cloud before the subsequent run, and the transformations were cumulatively composed. ICP was repeated five times for GeDi+ICP and 20 times for ICP alone.

### 3.2 Experimental Settings

#### 3.2.1 Dataset

The datasets consisted of jawbone CT volumes from seven patients diagnosed with jaw deformity at Otowa Hospital, together with the pseudo-IOS point clouds generated from them. Each CT volume comprised axial slices of 512 × 512 pixels with an in-plane resolution of 0.3906 mm and a slice interval of 1.0 mm. The pseudo-IOS point clouds and corresponding CT-derived point clouds were generated independently for the mandible and maxilla.

The datasets were collected with the approval of the Ethics Committee of Kyoto University Graduate School and Faculty of Medicine (approval number: R5312). Because the study used retrospectively collected CT imaging data and direct consent from all participants was not feasible, an opt-out approach was adopted in accordance with the ethical guidelines. An information disclosure document was made publicly available, providing participants with the opportunity to decline participation. No participant opted out of the study.

#### 3.2.2 Artifact conditions

For each case and jaw, nine artifact conditions were generated by designating $n_m = 0, 1, \cdots, 8$ teeth as metal regions. The $n_m = 0$ condition corresponded to the original artifact-free CT and was categorized as the without-artifact condition, whereas the $n_m = 1, \cdots, 8$ conditions were categorized as with-artifact conditions. For the two-level repeated-measures analysis, the registration metrics obtained for $n_m = 1, \cdots, 8$ were averaged within each case, jaw, and registration method, to yield one value for the with-artifact condition. Following Nakao et al. [3], eight teeth were selected in a predefined order, comprising two randomly selected posterior teeth, two adjacent posterior teeth, two anterior teeth, and two teeth randomly selected from the remaining dentition. The conditions with 1–8 metal teeth were generated by cumulatively adding the selected teeth in this order.

#### 3.2.3 Comparison of Registration Methods

We compared three registration conditions: (i) ICP alone, in which the perturbed source cloud is aligned to the target directly by ICP; (ii) GeDi alone, in which the global alignment is obtained from descriptor-based correspondence matching without subsequent refinement; and (iii) GeDi+ICP (proposed), in which the GeDi global alignment is refined by ICP. Comparing these three conditions isolates the contribution of the global (GeDi) and local-refinement (ICP) stages, clarifying the behavior of each under artifact-affected data.

#### 3.2.4 Evaluation Metrics

Because the pseudo-IOS point cloud is generated from the CT, within the same coordinate frame, the ground-truth transformation $g^*$ is known, and the correspondence between the estimated and ground-truth point sets is available at the level of individual points. We evaluated registration accuracy using the following three metrics and computation time. Mean absolute error (MAE) is the mean Euclidean distance between corresponding points of the estimated and ground-truth point clouds, computed over the known point correspondences. It directly reflects the deviation of the estimated alignment from the ground truth. Mean distance (MD) is the mean of the bidirectional nearest-neighbor distances between the registered source cloud and the target cloud. Unlike MAE, it does not rely on the known correspondence and instead measures surface-to-surface closeness. Hausdorff distance (HD) is the maximum of the nearest-neighbor distances between the two clouds and captures the largest local misalignment. The time required to complete the entire registration pipeline was measured for each method.

For statistical analysis, the registration metrics from the 10 trials were averaged for each case, jaw, artifact condition, and registration method. For the with-artifact category, the resulting values for ($n_m = 1, \cdots, 8$) were further averaged within each case, jaw, and method, yielding one value per case for each combination of method, artifact presence, and jaw. A three-way repeated-measures ANOVA (method × artifact × jaw, with case as the repeated factor) was then conducted on the MAE results for GeDi and GeDi+ICP.

## 3.3 Results

Table 1 shows the mean and standard deviation of the MAE, MD, and HD (mm) for each method under each combination of artifact condition (with/without metal artifacts) and jaw (mandible, maxilla), averaged over the seven cases. ICP alone produced very large errors under every condition (overall MAE 25.89 mm, HD 20.19 mm), indicating that, from the perturbed initial position, it frequently converged to local minima and failed to align the point clouds. Although GeDi+ICP was slower than the other methods, its runtime of approximately 10 s remains feasible for a registration workflow. We therefore treated ICP as a baseline that illustrates the difficulty of registration without a reliable initialization, and focused the statistical analysis on the comparison between GeDi and our proposed GeDi+ICP pipeline.

Figure 2 shows representative registration results without metal artifacts and with eight simulated metal teeth. ICP converged to anatomically incorrect local solutions: without artifacts, the source cloud aligned with an inverted superior–inferior orientation, whereas in the presence of artifacts, the mandibular source aligned to an incorrect region of the target. GeDi recovered the global position but retained visible residual discrepancies, particularly in the artifact-affected condition. Subsequent ICP refinement produced an alignment closest to the ground truth. These observations were consistent with the overall quantitative findings that GeDi provided global initialization, and ICP reduced the residual error.

### 3.3.1 Registration Accuracy

GeDi+ICP achieved the smallest error across all conditions, with MAE ranging from 0.55 to 0.69 mm and HD remaining near or below 1.1 mm. GeDi alone was less accurate (MAE 1.09–3.64 mm), and importantly, was markedly more sensitive to metal artifacts: under the artifact condition, its MAE for the mandible increased to 3.64 mm, whereas that for GeDi+ICP remained at 0.69 mm. GeDi+ICP also showed consistently smaller standard deviation (e.g., 0.06 mm vs. 1.16 mm for the artifact-affected mandible), indicating more stable registration. The same tendencies were observed for MD and HD.

### 3.3.2 Statistical Analysis

A three-way repeated-measures ANOVA showed significant main effects of method ($F(1,6) = 58.82$, $p < .01$), artifact ($F(1,6) = 35.62$, $p < .01$), and jaw ($F(1,6) = 20.44$, $p < .01$). The method × artifact ($F(1,6) = 27.10$, $p < .01$), method × jaw ($F(1,6) = 35.26$, $p < .01$), and artifact × jaw interactions ($F(1,6) = 30.59$, $p < .01$) were also significant, as was the method × artifact × jaw interaction ($F(1,6) = 27.41$, $p < .01$). The condition means indicated that these interactions were primarily associated with the marked artifact-related increase in GeDi error for the mandible,

**Table 1 Registration errors and computation time (mean ± SD) for each method and artifact condition. Mean absolute error (MAE), mean distance (MD), and Hausdorff distance (HD) are reported separately for the mandible and maxilla, whereas computation time is summarized across both jaws. Check marks denote the with-artifact conditions. Lower values indicate better performance; the best and second-best values within each condition are shown in bold and underlined, respectively. ICP, iterative closest point algorithm; GeDi, general and distinctive 3D local descriptors.**

| Method | artifact | MAE (mm) | | MD (mm) | | HD (mm) | | Time (s) |
|---|---|---|---|---|---|---|---|---|
| | | Mandible | Maxilla | Mandible | Maxilla | Mandible | Maxilla | |
| ICP | | 30.96 ± 9.18 | 24.41 ± 9.11 | 11.57 ± 5.15 | 9.79 ± 4.72 | 21.69 ± 8.95 | 19.22 ± 7.20 | <u>5.65 ± 1.12</u> |
| ICP | ✓ | 27.69 ± 10.16 | 20.52 ± 7.56 | 12.32 ± 5.38 | 9.57 ± 4.96 | 22.34 ± 8.43 | 17.52 ± 7.19 | **5.64 ± 1.06** |
| GeDi | | 1.17 ± 0.22 | 1.09 ± 0.24 | 0.48 ± 0.05 | 0.45 ± 0.04 | 2.01 ± 0.35 | 1.81 ± 0.37 | 8.89 ± 2.08 |
| GeDi | ✓ | 3.64 ± 1.16 | 1.61 ± 0.53 | 1.17 ± 0.36 | 0.57 ± 0.12 | 5.70 ± 1.66 | 2.65 ± 0.88 | 9.72 ± 2.63 |
| GeDi + ICP | | **0.55 ± 0.14** | **0.55 ± 0.08** | **0.23 ± 0.05** | **0.23 ± 0.04** | **0.84 ± 0.21** | **0.82 ± 0.16** | 9.14 ± 2.16 |
| GeDi + ICP | ✓ | <u>0.69 ± 0.06</u> | <u>0.64 ± 0.06</u> | <u>0.29 ± 0.02</u> | <u>0.26 ± 0.02</u> | <u>1.10 ± 0.09</u> | <u>0.96 ± 0.14</u> | 10.05 ± 2.78 |

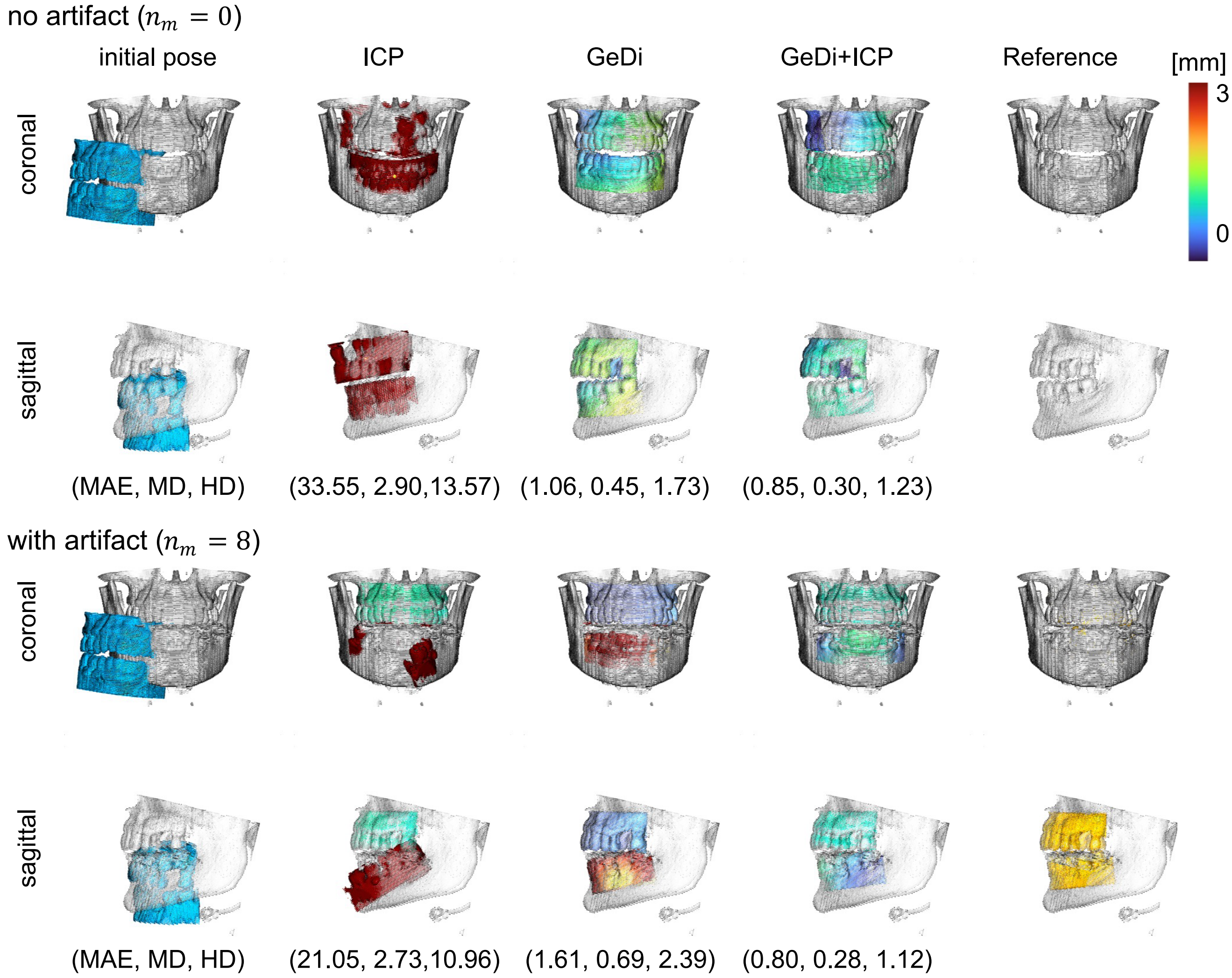

**Fig. 2 Registration results without metal artifacts and with eight simulated metal teeth. Columns show the initial position, ICP, GeDi, GeDi+ICP, and reference alignments. Coronal and sagittal views are presented for each condition. For ICP, GeDi, and GeDi+ICP, the transformed pseudo-IOS source points are color-coded according to their point-wise distances to the reference pseudo-IOS source. The same color scale is used across all panels, with distances expressed in millimeters. The values below each result represent MAE, MD, and HD in millimeters.**

whereas GeDi+ICP showed comparatively small changes across jaw and artifact conditions.

These results indicate that once ICP is provided with a reliable initialization from GeDi, the proposed GeDi+ICP attains significantly higher accuracy and lower variability than GeDi alone, and that this advantage is more pronounced in artifact-affected mandibles.

## 4. Discussion

The present study proposes a pseudo-IOS framework for evaluating CT–IOS registration against a known ground-truth transformation, a framework involving a coarse-to-fine method combining GeDi with ICP. GeDi+ICP achieved stable registration from perturbed initial positions, with an MAE of 0.55–0.69 mm across the jaw and artifact conditions, whereas ICP alone frequently converged to incorrect local solutions. The results indicate that descriptor-based global alignment and geometry-based local refinement can play complementary roles in this registration problem.

The contrast between ICP alone and GeDi+ICP highlights the importance of initialization in the registration. Because it minimizes local surface distances, ICP is effective only when the source is sufficiently close to the correct alignment; otherwise, it can converge on a local minimum. GeDi instead estimates the global position from local-descriptor correspondences without relying on the initial position. However, GeDi alone retained residual errors, particularly under the artifact-affected condition. Subsequent ICP refinement corrected these discrepancies once an approximately correct global position was obtained. Thus, the performance of the proposed method arises from combining two stages with different strengths, rather than from the inclusion of either stage alone.

Metal artifacts had a statistically significant effect on registration accuracy, but their absolute influence was substantially smaller for GeDi+ICP than for GeDi alone. This suggests that artifact-induced surface distortion primarily disrupts descriptor correspondences during global alignment, and that ICP can compensate for part of this error by exploiting the broader surface overlap available after initialization. Previous studies reported deterioration of markerless CT/CBCT–IOS registration in the presence of metallic or highly radiopaque restorations [14, 15].

The pseudo-IOS framework provides a controlled means for separating registration error from uncertainty in the reference alignment. Because the pseudo-IOS and CT-derived clouds are generated within the same coordinate system, both the transformation and point-level correspondences are known. This enables direct evaluation using MAE and surface-based metrics. Nevertheless, the framework does not reproduce the full domain gap between independently acquired CT and IOS data. The pseudo-IOS lacks scanner noise, scanning-path deformation, soft-tissue interference, and partial occlusion, while both clouds share the same underlying CT geometry.

Study limitations include the small dataset and the simulated rather than real metal artifacts. Although GeDi+ICP maintained submillimeter MAE across all evaluated conditions, the HD reached 1.10 mm, and these geometric results alone do not establish task-specific clinical acceptability. Validation using real CT–IOS pairs, larger cohorts, and real metal artifacts is therefore required.

## 5. Conclusion

We developed a pseudo-IOS framework that enables quantitative evaluation of CT–IOS registration against a known ground-truth transformation. GeDi-based global alignment followed by ICP refinement achieved submillimeter MAE across all evaluated jaw and artifact conditions, and was significantly more accurate than GeDi alone. Although metal artifacts affected registration accuracy, their absolute impact on GeDi+ICP was small.

### Conflict of interest

The authors declare no conflicts of interest with any companies or commercial organizations per the definition of Japanese Society for Medical and Biological Engineering.

### Ethics declaration

This study was conducted with the approval of the Ethics Committee of Kyoto University Graduate School and Faculty of Medicine (approval number: R5312). Because the study used retrospectively collected CT imaging data and direct consent from all participants was not feasible, an opt-out approach was adopted in accordance with the ethical guidelines. An information disclosure document was made publicly available, providing participants with the opportunity to decline participation. No participant opted out of the study.

### Acknowledgements

This work was supported by JSPS KAKENHI Grant Numbers JP24H00795 and JP24K22405. We thank Edanz (https://jp.edanz.com/ac) for editing a draft of this manuscript.